\documentclass[aps,pra,twocolumn,unsortedaddress]{revtex4-1}

\usepackage{graphicx}
 \usepackage{amsmath}
 \usepackage{multirow}

\begin{document}

\title{Scaling behaviour of the ground-state antihydrogen yield from CTMC simulation as a function of positron density and temperature}

\author{B. Radics}
\email{balint.radics@riken.jp}
\author{D.J. Murtagh}
\author{Y. Yamazaki}
\affiliation{Atomic Physics Laboratory, RIKEN, Saitama 351-0198, Japan}
\author{F. Robicheaux}
\affiliation{Department of Physics, Purdue University, West Lafayette, Indiana 47907, USA}


\begin{abstract}

Antihydrogen production has reached such a level that precision spectroscopic measurements of its properties are within reach. In particular, the ground-state level population is of central interest for experiments aiming at antihydrogen spectroscopy. The positron density and temperature dependence of the ground-state yield is a result of the interplay between recombination, collisional, and radiative processes. Considering the fact that antihydrogen atoms with the principal quantum number n=15 or lower quickly cascade down to the ground state within 1ms, the number of such states are adopted as a measure of useful antihydrogen atoms. 
It has been found that the scaling behaviour of the useful antihydrogen yield is different depending on the positron density and positron temperature.

\end{abstract}


\maketitle


\section{Introduction}

Antihydrogen, the simplest antiatom consisting of a positron and an antiproton, is expected to provide one of the most stringent test grounds of the CPT symmetry, the most fundamental symmetry in the standard model. For this purpose, its formation  \cite{WOelert} \cite{Amoretti_Nat2002} \cite{Gabrielse_PRL2002} and its trapping  \cite{Andresen_Nat2010} \cite{Alpha_NatPhys2011} \cite{Gabrielse_PRL2012}  have been the focus of several antimatter experiments in the last two decades. Very recently, the preparation of antihydrogen beam for in-flight spectroscopy has been successfully demonstrated \cite{Kuroda_NatComm2014}. The goal of the experiments is to prepare ground-state antihydrogen for precision measurements. The sensitivity of these measurements usually depends on the ground-state level population.

Antihydrogen is produced by mixing together cold antiprotons and positrons. It is well known that the dominant formation mechanism is three-body recombination to Rydberg states, while the subsequent evolution of level population is governed by collisional (de)excitation and ionisation. The list of the important processes is given in Table~\ref{proc_tab}. The positron temperature and density dependence of the rates of recombination and ionisation or (de)excitation processes are different. The rates of these collisional processes and antihydrogen formation have been previously studied using Classical Trajectory Monte Carlo (CTMC) methods \cite{Glinsky_1991}  \cite{Hurt_JPB2008} \cite{Pohl_2008} \cite{BassDubin_2009}. The first detailed analysis of the three-body recombination rate in a strongly magnetised, cryogenic plasma was performed by Glinsky and O'Neil \cite{Glinsky_1991}. Robicheaux and Hanson \cite{Robicheaux_2004} studied the three-body recombination rate and antihydrogen formation for positron temperature $4 \leq T_{e} \leq 16$ K and magnetic field strength $B = 3$ T and $5.4$ T. Another study of antihydrogen formation was done by Jonsell et al. \cite{Jonsell_2009} for the conditions of the ATHENA experiment \cite{Amoretti_Nat2002}, using positron temperature scale of $T_{e} = 15$ K and magnetic field strength of $B = 3$ T. A detailed study has been published by Bass and Dubin \cite{BassDubin_2009} using Monte Carlo simulations for antihydrogen formation in cryogenic temperatures and strongly magnetised environment. In their work Bass and Dubin studied antihydrogen formation at low positron temperature and high magnetic field, and characterised such a regime with a magnetisation parameter, $\chi \equiv \bar{v}/b\Omega_{c} = 0.0018(T_e/4\; K)^{3/2}/(B/6\;T)$, where $T_e$ is positron temperature, $B$ magnetic field strength, $\bar{v}$ positron thermal speed, $b$ classical distance of closest approach and $\Omega_{c}$ positron cyclotron frequency.  The magnetisation parameter explored in their work ranges from $\chi = 0$ (infinite magnetic field) to $\chi = 0.005$.  
The later value corresponds \emph{e.g.} to a combination of $T = 10$ K positron temperature and $B = 10$ T magnetic field strength.\\
\begin{table}[ht]
	\begin{tabular}{|l|l|} \hline \hline
	  \bf{Process name} & \bf{Process} \\ \hline 
	   Three-body recombination  & $\bar{p} + e^{+} + e^{+} \rightarrow \bar{H}^{*} + e^{+}$ \\ 
	   Radiative recombination & $\bar{p} + e^{+} \rightarrow \bar{H}^{*} + h\nu $ \\
	   Collisional (de)excitation & $\bar{H}^{*} + e^{+} \leftrightarrow \bar{H}^{**} + e^{+}$ \\
	   Collisional Ionisation & $\bar{H}^{*} + e^{+} \rightarrow \bar{p} + e^{+} + e^{+}$ \\
	   Spontaneous radiative decay & $\bar{H}^{**} \rightarrow \bar{H}^{*} + h\nu$ \\
	   Stimulated radiative transition & $\bar{H}^{*} + h\nu \rightarrow \bar{H}^{**} (+ 2 h\nu)$ \\
	   \hline \hline
	\end{tabular}  
	\caption{Atomic scattering processes during antihydrogen production. A star ($^{*}$) or two stars ($^{**}$) denote two different quantum states.}
	\label{proc_tab}
\end{table}
However, in some of the more recent experiments the positron temperature may be higher than studied before  \cite{Kuroda_NatComm2014} \cite{AmoleAlpha_2013}. In this work we study the magnetisation parameter range of $0.04 \leq \chi \leq 3.5$, consistently covering both strongly and weakly magnetised plasma conditions, unlike previous studies. The calculation contains three-body and radiative recombination, ionisation and collisional (de)excitation, spontaneous and stimulated radiative processes. \\

This paper is organised as follows. In section~\ref{sec:LevPopMod} the level population model is outlined. First the atomic population evolution equations are described, then the computation of the various rates is discussed. In section~\ref{sec:Discussion} the model is used to compute level populations for typical experimental plasma conditions. Finally, section~\ref{sec:Conclusion} concludes our findings.


\section{\label{sec:LevPopMod}Level population model}

\subsection{Evolution of atomic populations}
Antihydrogen is formed by mixing antiproton and positron plasmas. For the formation of bound states two processes are responsible: radiative recombination and three-body recombination. However, after bound states are formed many other processes can take place: collisional (de)excitation and ionisation by positron impact, spontaneous radiative decay and stimulated radiative transitions due to the presence of a finite temperature black-body radiation field. For a particular bound state, $N(i)$ (where $N$ is level population, \emph{i.e.} number of atoms in bound state $i$, where $i$ is any quantum number), the scattering processes either populate or depopulate a given state. Therefore the evolution of level population of each state and the loss of antiprotons over time can be described by a system of coupled differential equations. For the evolution of the level population of bound states ($i$ running over the quantum numbers chosen),
\begin{equation}
\label{eq:difflevpop}
\begin{split}
\frac{dN(i)}{dt} &= [C_{rr}(i) + C_{tbr}(i)n_{e}]n_{e}N_{p} - C_{ion}(i)n_{e}N(i) \\
& + \sum_{j \neq i}[C_{col}(j, i)n_{e} + C_{str}(j, i)]N(j) \\
& - N(i)\sum_{j\neq i}[C_{col}(i,j)n_{e} + C_{str}(i,j)],
\end{split}
\end{equation}
while the number of bare antiprotons fulfills the relation given by,
\begin{equation}
\frac{dN_{p}}{dt} = \sum_{i} \left( C_{ion}(i)n_{e}N(i) - [C_{rr}(i) + C_{tbr}(i)n_{e}]n_{e}N_{p} \right) .
\end{equation}
Here $N_{p}$ is the number of antiprotons, $n_e$ is the density of positrons, $C_x$ denote rate coefficients for atomic process, $C_{rr}(i)$ and $C_{tbr}(i)$ denote rate coefficients for radiative and three-body recombination to state $i$ respectively, $C_{ion}(i)$ denotes ionisation by positron impact from bound state $i$, $C_{col}(i,j)$ denotes collisional excitation or deexcitation by positron impact from state $i$ to state $j$ and $C_{str}(i,j)$ denotes spontaneous or stimulated transitions due to presence of a radiation field.

The coupled set of differential equations are solved using Bulirsch-Stoer method \cite{GSL}. The dimensionality of the problem depends on the number of bound states used during computation. The initial conditions used for the system is a completely empty antihydrogen level population and a given number of antiprotons. During the evolution antihydrogen bound states are formed by filling the empty level population according to equation~\eqref{eq:difflevpop}.


\subsection{Computation of rates of atomic processes}
The rate coefficient of an atomic scattering process depends on the energy distribution of the impact particle, $C = \int \sigma(E)f(E) v dE$, where $f(E)dE=2\sqrt{E/\pi}(k_B T_{e})^{-3/2}\exp(-E/k_{B}T_{e})dE$ is the energy distribution function with positron temperature $T_{e}$, $\sigma(E)$ is the cross section as a function of energy, $E$, $v$ is the velocity of the impact particle and $k_B$ is the Boltzmann constant. The rate coefficient can be computed by Monte Carlo methods, randomly sampling the energy distribution of the projectile ($e^{+}$), counting the occurrence of various final states and scanning with the impact parameters used for the orbitals. The collision can be simulated by integrating the coupled Newton's equations of motion for the projectile and bound positron, and using a fixed antiproton position. In case of magnetic field given in a certain direction the coordinates can be chosen such that the integration is done in a cylindrical system and the initial microcanonical ensemble distributions can be prepared accordingly \cite{Robicheaux_PRA2006}. From the computed rate of ionisation the three-body recombination rate can be derived using the detailed balance principal and assuming thermodynamical equilibrium
\begin{equation}
C_{tbr}(i) = C_{ion}(i)g_{i}n_{e}\Lambda^{3}e^{\frac{E(i)}{k_{B}T_e}},
\end{equation}
where $g_{i} = n_{i}^{2}$ is the statistical weight of state $i$ with principal quantum number $n_{i}$, $\Lambda=h/\sqrt{2\pi m_e k_{B} T_{e}}$ is the thermal de Broglie wavelength of the positron, $m_{e}$ is the positron mass, $E(i)$ is the binding energy of state $i$. In the computation the detailed balance principal is also applied between the collisional excitation and de-excitation rates.

The radiative recombination rate is calculated as follows. For each $n$ principal and $l$ orbital quantum number, the radial part of the wave function is computed using a finite difference method based on the Numerov algorithm. Then the energy normalised continuum state for angular momenta $l-1$ and $l+1$ is computed at energy $E$ using the same finite difference algorithm. The recombination rate to a specific $n, l$ in atomic units is then
\begin{equation}
\begin{split}
&\langle \sigma_{rr} v \rangle_{n,l} = \left(\frac{2\pi}{T_{e}}\right)^{3/2} \left(\frac{4}{3}\right) \alpha^{3} \times \\
& \times \int dE (E-E_{nl})^{3} e^{-\frac{E}{k_{B}T_{e}}} \left[(l-1) D^{2}_{E,l-1} + l D^{2}_{E, l+1} \right] ,
\end{split}
\end{equation}
where $\alpha$ is the fine structure constant, $D_{E,l}$ denotes dipole matrix elements calculated for energy $E$ and orbital quantum number $l$. The dipole matrix elements are computed using numerical integration over radial coordinate $r$, and cross-checked by recomputing using length and velocity gauge.

Spontaneous radiative decay rates are calculated from Einstein A-coefficients for hydrogen orbitals. Stimulated transition rates for an external black-body radiation field, with temperature $T_{r}$, are added using the radiative transition rates as
\begin{equation}
C_{em}(i,j) = C_{rad}(i,j)\left[1 + \frac{1}{\exp(E_{ij}/k_{B}T_{r})-1}\right], i > j,
\end{equation}
\begin{equation}
C_{abs}(i,j) = C_{rad}(j,i)\frac{1}{\exp(E_{ij}/k_{B}T_{r})-1}, i < j,
\end{equation}
where $C_{rad}(i,j)$ is the spontaneous radiative transition rate, $C_{em}(i,j)$ and $C_{abs}(i,j)$ are the stimulated emission and absorption rates  between states $i$ and $j$, $E_{ij} = |E(i) - E(j)|$ are the binding energy differences.

Rate coefficients have been calculated for various positron temperatures and magnetic field strengths. For zero magnetic field our calculation agrees with result of Pohl \emph{et al.} \cite{Pohl_2008}. Thermodynamical equilibrium detailed balance was applied between the rates of direct and indirect processes for collisional (de)excitation, such that the following relation is established
\begin{equation}
\frac{C(n_{i}\rightarrow n_{f})}{C(n_{f}\rightarrow n_{i})} = \frac{n_{f}^{2}}{n_{i}^{2}}e^{\frac{E(f)-E(i)}{k_{B}T_{e}}} 
\end{equation}
where $C(n_{i}\rightarrow n_{f})$ denotes rate of a process from initial state $n_i$ to final state $n_f$.


\section{\label{sec:Discussion}Discussion}

\subsection{Level population evolution for experimental conditions}

\begin{figure}
\includegraphics[width=8.6cm]{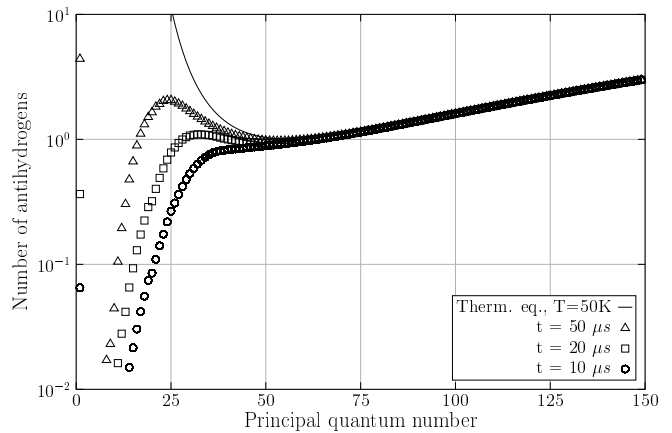}
\caption{Antihydrogen bound-state level population distribution after evolution of 10 $\mu$s (circle), 20 $\mu$s (square) and 50 $\mu$s (triangle) and the thermal equilibrium level population distribution (solid line) for positron temperature of $T_e = 50 K$ and positron density $n_{e} = 10^{14}$ m$^{-3}$.}
\label{levpop_plot1}
\end{figure}

The coupled system of differential equations are solved for example: 150 bound states, corresponding to principal quantum numbers $n_{i} = 1\dots 150$, a number chosen sufficiently large to represent thermal equilibrium level population for highly excited Rydberg states.  The initial conditions for the solution are an empty level population for the bound states of antihydrogens, $N_p = 10^{6}$ antiprotons, a positron density of $n_e = 10^{14}$ m$^{-3}$, a positron temperature of $T_e = 50$ K, and a magnetic field strength of $B = 2$ T. A typical result is shown in Figure~\ref{levpop_plot1} for a range of evolution times 10 $\mu$s (circle), 20 $\mu$s (square) and 50 $\mu$s (triangle).  Also, the thermal equilibrium distribution of bound state population, given by the Saha-Boltzmann relation, is shown for the same temperature of $T_{e} = 50$ K (solid line),
\begin{equation}
N_{th}(i) = N_{p} n_{e} n_{i}^{2} \Lambda^{3} e^{\frac{E(i)}{k_{B}T_{e}}}.
\end{equation}

A typical experimental time scale for evolution is 10 $\mu s$ (empty circles in Figure~\ref{levpop_plot1}), which is the average time an antiproton spends in the positron plasma. This timescale is set by the typical thermal speed of antiprotons in these experiments at $T_{e}= 300$ K temperature and the typical length scale of the positron plasma of $L_{e^{+}} \simeq 20$ mm \cite{Kuroda_NatComm2014}. 
For highly excited Rydberg states (principal quantum number, $n_{i} \geq 50$) after a few $\mu s$ of evolution time, a thermal equilibrium  is established, while for deeper states the system does not reach thermal equilibrium. The rates of the various processes shaping the level population distribution change differently as a function of the principal quantum number, $n_{i}$: radiative decay rates decrease rapidly with $n_i$, positron-Rydberg atom collision rates increase with $n_{i}$, capture by radiative recombination favours low $n_i$, while capture by three-body recombination favours high $n_i$. There is a characteristic excited state level below which the level population is depleted, eventually reaching a point where the spontaneous radiative decay rates dominate and ionisation rate becomes very low. 

\begin{figure}
\includegraphics[width=8.6cm]{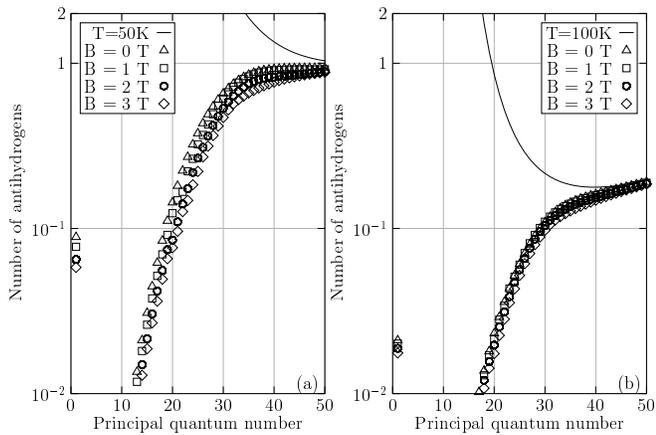}
\caption{Antihydrogen bound-state level population distribution after evolution of 10 $\mu s$ for various magnetic field strength  $0 \leq B \leq 3 $T, and the thermal equilibrium level population distribution (solid line) for positron temperatures $T_e = 50 $K (a) and $T_e = 100 $K (b) and positron density $n_{e} = 10^{14}$ m$^{-3}$.}
\label{levpop_plot1B}
\end{figure}

In Figure~\ref{levpop_plot1B}  the impact of variation of magnetic field on the level population is shown for $T_e=50$ K and $T_e=100$ K positron temperatures. While at $T_e = 50$ K the ground-state level population is reduced by $\sim 50 \%$ with increasing magnetic field strength, already at $T_{e} = 100$ K the level population does not reduce significantly even at high magnetic field strengths. In the following discussion the magnetic field strength is fixed at $B= 2$ T.

\subsection{Positron plasma parameter scan}

\begin{figure}[ht]
\includegraphics[width=8.6cm]{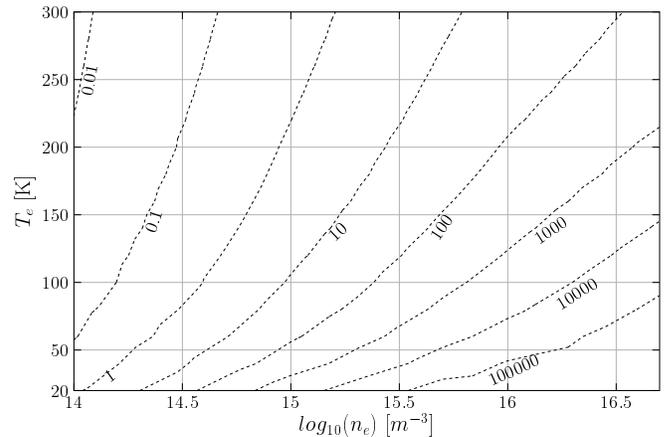}
\caption{Contour plot of the number of ground-state antihydrogen atoms for various positron temperature (vertical axis) and density values (horizontal axis).}
\label{maptempdens_iso}
\end{figure}

The aim of this work is to investigate the evolution and scaling behaviour of the ground-state level population of antihydrogen atoms with a wide range of positron plasma parameters. Given that the evolution is governed by a large set of coupled differential equations, in which the population of a particular quantum state could receive contribution from any other quantum state, the scaling behaviour of the ground-state level population  cannot be calculated with a simple formula. For each positron plasma density and temperature the conditions at the beginning and during evolution are different, therefore for each case the evolution will also take a different path.  In order to study the effect of different plasma conditions on the ground-state level population a parameter scan of positron density and temperature has been performed in the range $10^{14} \leq n_e \leq 10^{16} m^{-3}$ and $20 \leq T_{e} \leq 300$ K. A new level population simulation was carried out for each temperature and density point, using $N_p = 10^{6}$ initial antiprotons, with a fixed evolution time of 10 $\mu$s. \\
For experiments attempting ground-state antihydrogen spectroscopy the figure of merit is the number of ground-state antihydrogen atoms with respect to plasma parameters. Experimentally there is usually a defined trapping time or a flight-time after which the measurements are performed. During this time some fraction of the excited states will decay and populate the ground-state. Thus, the total ground-state yield may be defined as an integral of the level population from $n=1$ to some upper principal quantum number, which depends on the particular experiment. As a possible upper limit in the current work the $n=15$ level is used because the lifetime of direct transition from this level to ground-state is within an order of a millisecond, a typical timescale for an in-flight type experiment.\\
Figure~\ref{maptempdens_iso} illustrates the result of the plasma parameter scan for ground-state level population scaling behaviour, represented as a contour plot where each contour line corresponds to the same ground-state antihydrogen yield. It is generally expected that at low positron temperatures and high densities the antihydrogen yield is larger than at high temperatures and low densities. However, the rate of increase in the number of antihydrogens does not follow any simple scaling law. For example at $T_{e} = 40$ K for an order of magnitude increase in positron density the antihydrogen number increases by three orders of magnitude. However, at $T_{e} = 300$ K the number of antihydrogens increases only one to two orders of magnitude for the same increase of positron density. These findings confirm that the evolution of level population of antihydrogen atoms strongly depends on the positron plasma conditions at the start and during the evolution.
\begin{figure}[ht]
\includegraphics[width=8.6cm]{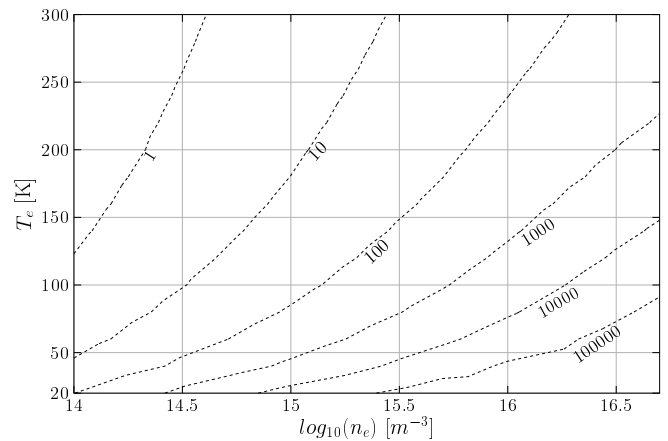}
\caption{Contour plot of the number of ground-state antihydrogen atoms for various positron temperature (vertical axis) and density values (horizontal axis) after 10 us interaction and 1 ms of flight.}
\label{maptempdens_iso_Flight}
\end{figure}

Once the antihydrogen leaves the positron plasma they will no longer undergo collisional or recombination processes. Only spontaneous radiative decay and black-body radiation govern the level population rate equations. The level population evolution during a 1 ms flight has been simulated by integrating the same rate equations (eq. (1)) but only considering the radiative processes and setting the initial value of the level population simulation according to the 10 $\mu s$ simulation result shown in Figure~\ref{maptempdens_iso}. The result is shown in Figure~\ref{maptempdens_iso_Flight}. The general trends for different positron density scaling across the temperature range remains, but the scaling behaviour seen in Figure~\ref{maptempdens_iso} is somewhat reduced by the radiative processes. At low temperatures an order of magnitude increase in positron density yields approximately two orders of magnitude increase in ground-state atoms, while at high temperatures the number of ground-state atoms increase only one order of magnitude. However, at very high positron density radiative decay does not seem to significantly impact the ground-state atom yield, because almost all antihydrogen atoms are in $n_i \leq 15$ states.

\begin{figure}[ht]
\vspace{20pt}
\includegraphics[width=8.6cm]{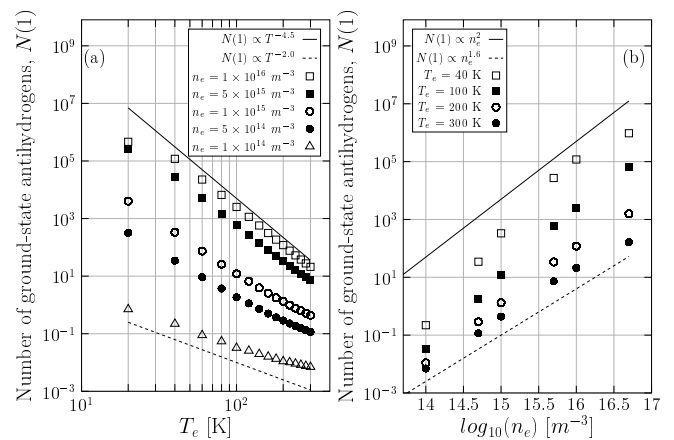}
\caption{Dependence of ground-state antihydrogen atoms on positron temperature (a) and density (b) for various positron density and temperature values (respectively) after 10 $\mu$s of mixing. The $\propto n_e^{2}T_{e}^{-4.5}$ (solid line) and $\propto n_{e}^{1.6}T^{-2.0}$ (dashed line) scaling behaviours are indicated for reference.}
\label{curvetempdens_OL}
\end{figure}

\begin{figure}[ht]
\vspace{20pt}
\includegraphics[width=8.6cm]{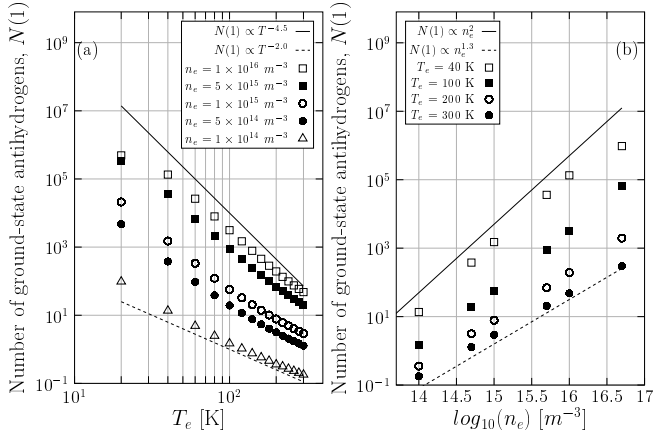}
\caption{Dependence of ground-state antihydrogen atoms on positron temperature (a) and density (b) for various positron density and temperature values (respectively) after 1 ms of flight. The $\propto n_e^{2}T_{e}^{-4.5}$ (solid line) and $\propto n_{e}^{1.3}T^{-2.0}$ (dashed line) scaling behaviours are indicated for reference.}
\label{curvetempdens_Flight}
\end{figure}

The power law scaling dependence of the number of ground-state antihydrogen atoms, $N(1)$, on positron-temperature and -density is shown in Figure~\ref{curvetempdens_OL} for 10 $\mu s$ of plasma interaction and in Figure~\ref{curvetempdens_Flight} for additional $1$ ms of flight. Also, in Figures~\ref{curvetempdens_OL} and \ref{curvetempdens_Flight}, the expected scaling behaviour is shown for three-body recombination (solid line), $R_{tbr} \propto n_{e}^{2}T_{e}^{-4.5}$, as well as a lower limit scaling which is fitted to the shape dependence found in this work (dashed line). 

In the case of the 10 $\mu s$ interaction (Figure~\ref{curvetempdens_OL}),  at \emph{e.g.} $T_{e} = 40$ K the ground-state antihydrogen atom number seems to increase with positron density as $N(1) \propto n_e^{2.5}$. However, at room temperature the positron density dependence drops to, $N(1) \propto n_e^{1.6}$.  At the same time the temperature dependence seems to deviate most from pure three-body-like behaviour at $n_e = 10^{14}$ m$^{-3}$ positron density, showing a positron temperature dependence $N(1) \propto T_e^{-2.0}$. While at positron density $n_e = 10^{16}$ m$^{-3}$ the temperature dependence becomes $N(1) \propto T_e^{-4.2}$. 
The behaviour of the power law scaling dependence suggests that at low positron temperature and high positron density the three-body recombination dominates the level population evolution, while at high positron temperature and low positron density collisional (de)excitation also becomes important during the evolution. In between these plasma conditions there is a transition region where the three-body recombination competes with the collisional deexcitation and the power law scaling behaviour of the ground-state level population deviates from that of both of these processes.

In case of an additional 1 ms of flight (Figure~\ref{curvetempdens_Flight}) the deviations from pure three-body recombination are similar in nature. The behaviour becomes closest to pure three-body recombination at low positron temperatures and high densities, 
while at higher temperatures or lower densities, the dependences on the density and temperature are both weaker.

It is interesting to note also that there is significant increase in the number of ground-state antihydrogen atoms at $n_e = 10^{14}$ m$^{-3}$ positron density. Both with and without 1 ms of flight $N(1)$ increases around two orders of magnitude when the positron temperature is reduced from $T_e = 300$ K to $T_e = 20$ K. \\
As a cross-check of our findings, in order to see which process dominates the evolution to the ground-state, we have performed the same parameter scan but turned off either radiative recombination or the collisional (de)excitation and ionisation processes. When radiative recombination is removed from the list of processes the impact on the contour plot in comparison with Figure~\ref{maptempdens_iso} is almost negligible. However, when all collisional processes are removed the ground-state yield is reduced such that only one ground-state atom is formed even at high positron densities ($n_{e} = 10^{16} m^{-3}$). 
This behaviour confirms that the evolution is an interplay of collisional (de)excitation and three-body recombination, with different relative dominance in various plasma conditions.\\

The simulation results indicate that the number of antihydrogen atoms determined experimentally does not follow simple power-law scaling. For example, our simulation predicts that a measurement of the ground-state antihydrogen atom yield as a function of the positron plasma temperature at a positron plasma density $n_e \simeq 10^{14}$ m$^{-3}$ would show a weaker power-law scaling behaviour than that of three-body recombination. It is noted that the positron temperature dependence of antihydrogen formation was measured at positron density $n_e \simeq 10^{14}$ m$^{-3}$, and an observed scaling behaviour of $T_{e}^{-1.1\pm 0.5}$ was reported \cite{Fujiwara_2008} by the ATHENA collaboration, which is consistent with the direction of our prediction qualitatively.


\section{\label{sec:Conclusion}Conclusion}

In this work ground-state antihydrogen formation and scaling behaviour was studied during mixing of antiprotons and positrons, using recent and potential experimental positron plasma conditions. The positron plasma magnetisation parameter range $0.04 \leq \chi \leq 3.5$ was explored using CTMC simulation combined with level population evolution. It was found that the scaling of the ground-state antihydrogen atoms does not follow a simple power-law behaviour, but it depends on the particular positron plasma conditions. The  positron plasma density and temperature determines the relative dominance between collisional (de)excitation and three-body recombination during the evolution. The results show that the ground-state antihydrogen yield deviates significantly from pure three-body recombination behaviour at higher positron temperature and lower density, and suggests that in this plasma condition collisional deexcitation becomes more important for ground-state population. In the case of low positron temperature and high density the three-body recombination behaviour is recovered, suggesting the importance of three-body process over collisional deexcitation to populate the ground-state level in this positron plasma condition. In the case of an additional 1 ms flight of the antihydrogen atoms, without recombination or collisions, the qualitative scaling behaviour remains similar, but it shows weaker dependence on positron density and temperature.  \\
\begin{acknowledgments}

We gratefully acknowledge conversations with J. Martin Laming, Yuri Ralchenko, and technical help of Imre Szeber\'enyi. This work was supported by the Grant-in-Aid for Specially Promoted Research (no. 24000008) of the Japanese Ministry of Education, Culture, Sports, Science and Technology (Monbukakagu-sho), Special Research Projects for Basic Science of RIKEN, RIKEN programme for young scientists, and the US National Science Foundation. Computational work was partially carried out at the Superman cluster of the Budapest University of Technology and Economics.  

\end{acknowledgments}

\bibliography{Bibliography}

\end{document}